\documentclass[useAMS,usenatbib]{mn2e}
\usepackage{graphicx}

\usepackage{amsmath}
\usepackage{amssymb}

\title[Filaments to Cores]{From Filaments to Oscillating Starless Cores}
\author[Keto, Burkert]{Eric Keto$^{1}$,\thanks{E-mail:
eketo@cfa.harvard.edu (EK); \hfill\break
burkert@usm.uni-muenchen.de (AB)} Andreas Burkert$^{2}$
\\
$^{1}$Harvard-Smithsonian Center for Astrophysics, 160 Garden St, Cambridge, MA 02420, USA \\
$^{2}$Ludwig-Maximilians-Universitat Muenchen
Universitat Sternwarte
Scheinerstr. 1
D-81679 Muenchen
}
\begin{document}

\date{Feb 6, 2013}


\maketitle

\label{firstpage}

\begin{abstract}
Long wavelength sonic oscillations are observed or inferred in many of the small, dark molecular clouds, the starless cores, that are the precursors to protostars. The oscillations provide significant internal energy and the time scale for their dissipation may control the rate of star formation in starless cores. 
Despite their importance, their origin is unknown. We explore one hypothesis
that the oscillations develop as a  starless core forms from a filament. We model this
process with a numerical hydrodynamic simulation and generate synthetic 
molecular line observations with a radiative transfer simulation. Comparison with
actual observations shows general agreement suggesting the proposed
mechanism is viable.

\end{abstract}

\section{Introduction}

In this paper we explore an hypothesis that the kinetic energy associated with 
the sonic oscillations  observed in starless cores comes from the gravitational 
energy released as a core forms from a filament by gravitational contraction.

The starless cores are dense concentrations of a few solar masses of molecular gas
and are
commonly observed in low-mass star-forming regions such as Taurus and
Ophiucus \citep{DiFrancesco2007,Andre2013}. 
They are significant as the future birthplaces of stars of  one-half to 
a few solar masses, the most  common stars in the galaxy
\citep{MyersBenson1983, Beichman1986}.
Their internal oscillations are significant because they may control the evolution
toward star formation. 
In star-forming regions such as Taurus, many starless cores
are massive and dense enough that their internal thermal energy provides only about
70\% of the energy needed for stability against gravitational collapse 
\citep{DickmanClemens1983,MyersBenson1983,Evans2001,Kandori2005, Kirk2005}. 
The remaining energy required for stability 
may be in the kinetic energy of the 
oscillations \citep{Keto2006,Broderick2010}. 

If oscillations support the cores, then the dissipation of the oscillations by
mode-mode coupling and radiation should result in gravitational collapse and
star formation if the oscillation energy is not resupplied \citep{Broderick2010}. 
In this case, the core lifetimes would be equal to the dissipation time \citep{Broderick2008}. 
This deduction is supported by 
some observations that suggest that more evolved cores, those with newly formed stars, 
have less turbulent energy than the starless cores \citep{Tachihara2002}.

Statistical arguments based on observation suggest that most cores have lifetimes
longer than their free-fall times \citep{Beichman1986,Jessop2000,Ward-Thompson2007}.
If many cores require additional energy for stability, and the oscillations are the source
of that energy, then the core lifetimes suggest that many cores are oscillating, not just the
few observed with complex modes \citep{Lada2003,Redman2006,Aguti2007}.

Despite the possible significance of the oscillations in the evolution of star-forming cores, 
the origin of the oscillations and the source of their energy is unknown. \citet{KetoField2005}
supposed that the oscillations were the expression of the subsonic turbulence at the
bottom of the supersonic turbulent cascade 
that characterizes the velocities in the larger-scale interstellar medium \citep{FBK2008}. However,
they did not specify further. Similarly, \citet{Broderick2010} assumed an initial turbulent
spectrum, either Kolmogoroff or flat, but again without specfiying the origin. 

The origin of the oscillations may be in the
process of core formation from filaments.  As a filament 
fragments and contracts
to form a spherical core, the gravitational energy released is converted into 
kinetic energy in oscillations. This hypothesis is motivated by both observation
and theory. Observations of low-mass star-forming regions
show that the cores are often associated
with filaments \citep{Goldsmith2008,Myers2009}. 
Theoretically, we know that self-gravitating
filaments are unstable to fragmentation and the formation of cores.  
It may be that
most cores form out of filaments.

We model the process of core formation from the contraction of a filamentary fragment
with a hydrodynamic simulation. We follow the evolution and at various time
we simulate observable spectral lines that can be compared with actual observations.
We find that in the initial contraction, the velocities
are mostly linear, along the filament, in the direction of the contraction. As the gas
rebounds off its own thermal pressure, the fragment flattens and expands perpendicular
to the length of the initial filament.  This is followed by a long period of
oscillations of slowly decreasing amplitude. 
Eventually the velocities become more and more radial
as the fragment becomes more and more spherical.

The profiles of the simulated spectral lines are
generally consistent with those observed. Our conclusion is that this is a plausible
hypothesis for the generation of the oscillations in starless cores.

\section{Numerical hydrodynamic simulation}

We use the 
smoothed-particle hydrodynamic code of  \citet{Wetzstein2009} and
follow the method of simulation as described in \citet{BurkertAlves2009}.
We begin the evolution with a finite cylinder to represent a fragment of
a filament. We set the mass of the filamentary 
fragment to 3 M$_\odot$ and
the sound speed of the molecular gas to 
0.2 km s$^{-1}$ corresponding to a gas temperature of 12 K. 

We know that if the fragment is bounded by an external pressure
it will evolve to a sphere in hydrostatic equilibrium, a Bonnor-Ebert (BE) sphere
\citep{Bonnor1956}. We want the final configuration to be stable against gravitational
collapse so we choose
the external pressure, which is constant throughout the simulation, to be just below
the pressure for critical stability of a sphere that is the same mass and isothermal
temperature as the initial cylindrical fragment. 
More precisely, we set the 
external pressure to the critical value for the stability of a (BE) sphere slightly more
massive (3.4 M$_\odot$) than our filamentary fragment.
In the simulation, confinement by an external pressure is 
simulated as, $P = \rho/c_s^2 -P_{ext}$ applied to all particles 
\citep[see ][]{BurkertAlves2009}.

With the mass, sound speed, and external pressure defined, it remains to
set the initial configuration of the filamentary fragment. 
The hydrostatic configuration for a pressure-bounded cylinder of infinite 
length \citep{Ostriker1964,FischeraMartin2012}  
is not completely appropriate because our filament has
a finite length. We therefore start
with a filament with a uniform density $1 \times 10^{-20}$  g cm$^{-3}$
and let it relax in the radial direction suppressing all
gas motion along the long axis (X-axis). The filamentary fragment 
evolves to a hydrostatic solution that is
taken as the initial condition for the subsequent simulation. 
The axial ratio of the filamentary fragment is 3:1.

The SPH simulations were run with 80K equal-mass particles, fulfilling the 
resolution requirement of 
\citet{BateBurkert1997}. The SPH smoothing length should be a fraction of the
dynamical and structural size scales.
It depends on the local density and in our clouds
is of order 5\% of the mean radius. The dominant oscillations are
on length scales that are a factor of 5 to 10 larger and are therefore well resolved.
The resolution was also tested by comparison with
a simulation of half as many particles.

\section{Radiative transfer}

We use the 3D non-LTE radiative transfer code MOLLIE \citep{Keto1990,KetoRybicki2010}
to produce simulated spectral lines as would be seen in a mm-wavelength radio observation.
The abundance of the molecular species is determined with a simple chemical model as
described in \citet{KetoCaselli2008,KetoCaselli2010} that includes the effects of freeze-out
at high density in the core or filament center and photodissociation near the boundary.
We produced radiative transfer 
simulations for CO, CS, HCO$^+$, and N$_2$H$^+$. Owing to the low mass
and low optical depth of the core, the spectral signatures of all the molecules were similar
despite their differing optical depths and critical densities for collisional de-excitation.
We show only the results for HCO$^+$ in the figures in this paper.

\section{Results}

\begin{figure*}
$
\begin{array}{cc}
\includegraphics[trim=0.10in 5.0in 0.8in 0.5in, clip,width=3.25in]{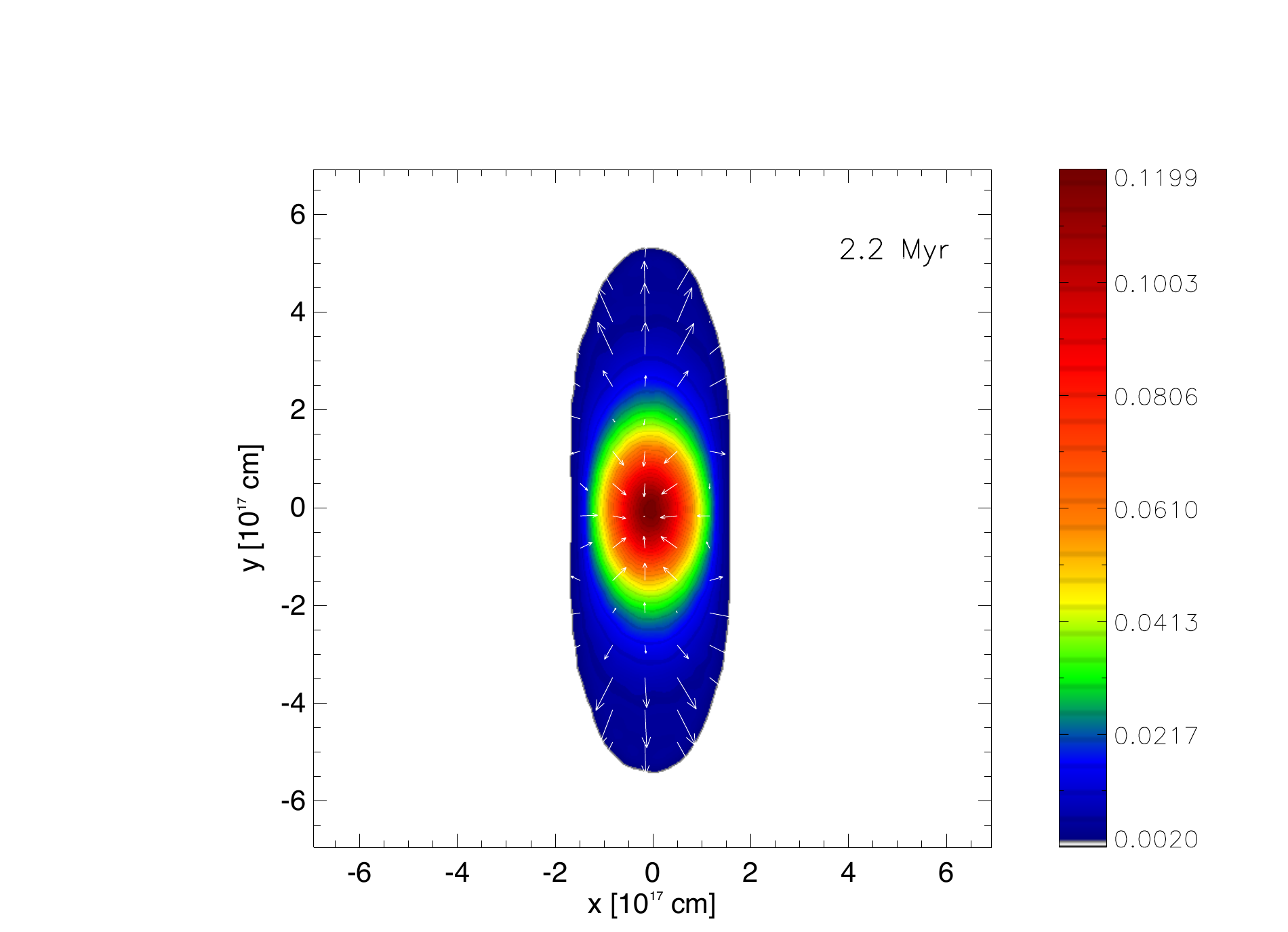}
\includegraphics[trim=0.10in 5.0in 0.8in 0.5in, clip,width=3.25in]{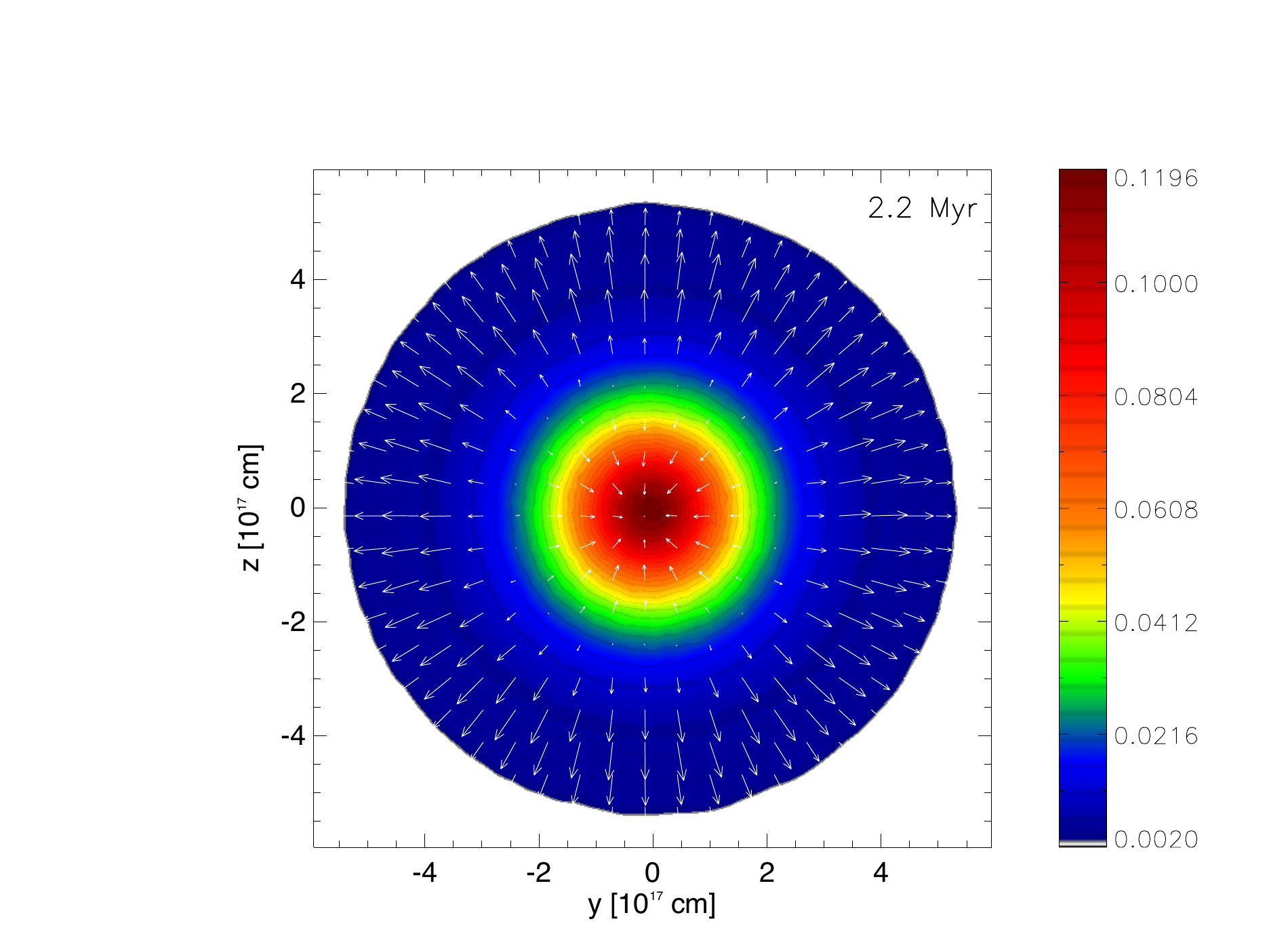}
\end{array}
$
\caption{
Densities and velocities on slices through two orientations at
2.2 Myr, ({\it left}) in the XY plane at Z=0, and ({\it right}) in the YZ planet X=0. The filament
was originally aligned with the X-axis. The axes are labeled in units of
$10^{17}$ cm and the density in units of $2\times 10^{-18}$ g cm$^{-3}$. The longest
arrow, the largest velocity is 0.62 km s$^{-1}$.
}
\label{fig:figxy-2p159}
\end{figure*}

The initial state of the filament is out of equilibrium longitudinally because of
the truncation to a fragment.  
The momentum 
of the initial longitudinal contraction along X causes the filament to flatten perpendicular to X
and expand in a disk shape along the YZ plane. 
Figure \ref{fig:figxy-2p159}  shows the  XY and YZ projections of
the fragment at this time (2.2 Myr).
The outer part of
the flattened fragment is expanding, rebounding from the longitudinal contraction 
while the inner part is radially contracting. 

The two perpendicular motions, longitudinal and radial, continue to oscillate between 
contraction and expansion with 
different periods, creating complex patterns of velocities and densities 
(figures \ref{fig:figxy-2p867} through \ref{fig:3p448_90_00_hcoplus}) even though the
initial state is highly symmetric.  
For example, in figure \ref{fig:figxy-2p867},
the gas density is highest in a ring around the core center.
The brightness of the spectral lines in figure 
\ref{fig:2p867_90_00_hcoplus} is correspondingly lower in the center.
This  shows how oscillations can create
peaks in the column density that are located away from the core center.
Off-center column density peaks are commonly seen in the 
simulations of fully turbulent cores 
\citep{Broderick2010}.

The simulation shows continuous damping of the velocities or amplitudes of the
oscillations. Figure \ref{fig:evolution} shows the average velocity defined by the 
total kinetic energy, $E_{kin}$  divided by the total mass, $M$,  
$v=\sqrt{2*E_{kin}/M}$.
The average velocity decreases by a factor of 3 
over a time scale of 10 Myr. In late phases, the typical oscillation velocity
is of order 0.03 - 0.05  km/s which is of order 20-30\% 
the sound speed.
The principle damping mechanism is by non-linear coupling
between the modes which is a 
decrease in momentum caused by collisions between fluid streams. In comparison
to the wave damping time, the free-fall and
crossing times are approximately 0.5 Myr. The figure also shows
the maximum density in the fragment which occurs in each cycle at the point
of maximum compression when the velocities are low and reversing. 
The effect of the two different periods, longitudinal and radial, is seen
in this figure. An animation of the oscillating filament may be seen at
http://www.usm.uni-muenchen.de/people/burkert/filament/index.html.

\begin{figure*}
$
\begin{array}{cc}
\includegraphics[trim=0.10in 5.5in 0.8in 1.0in, clip,width=3.25in,angle=180]{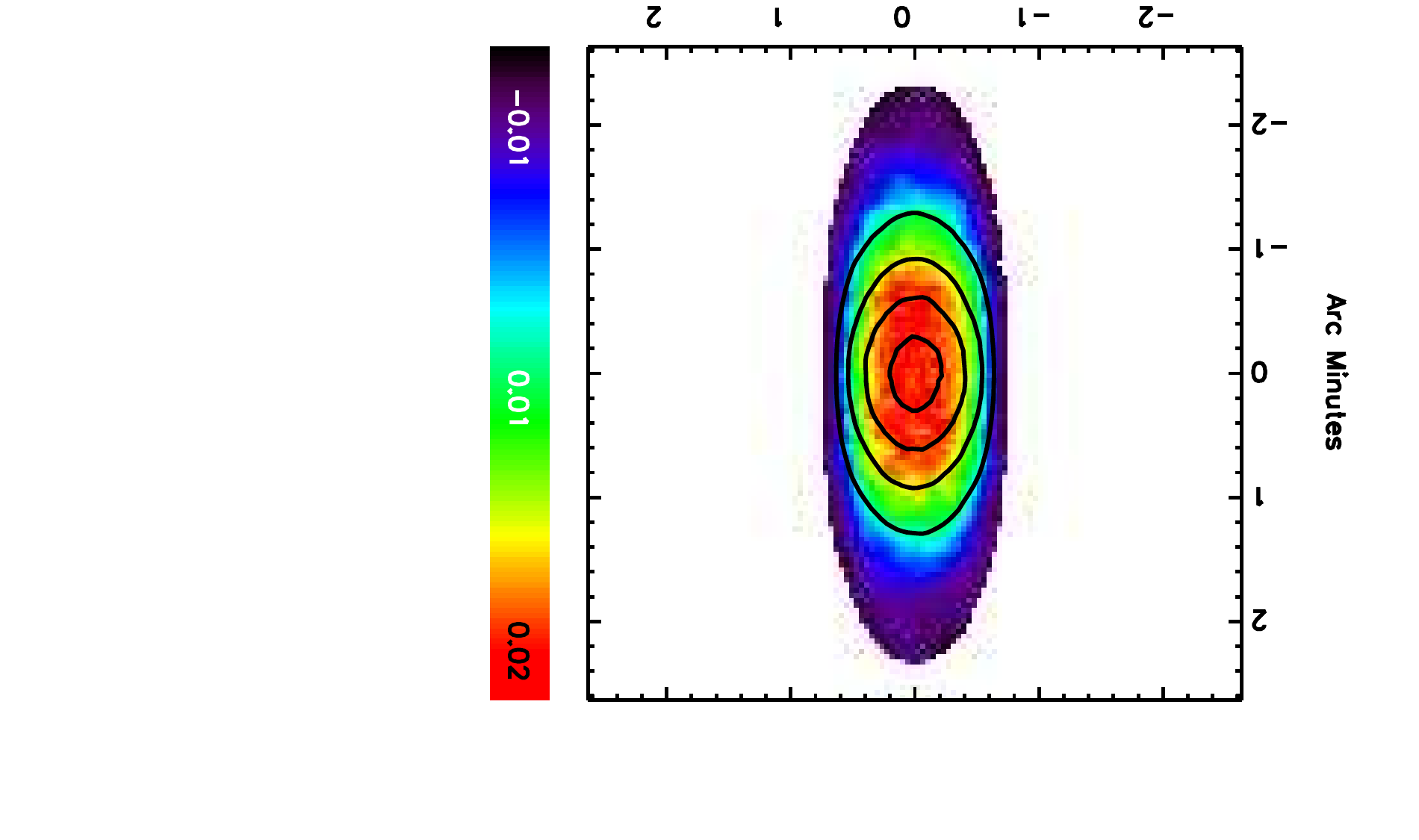}
\includegraphics[trim=0.10in 1.6in 0.8in 0.4in, clip,width=3.25in,angle=180]{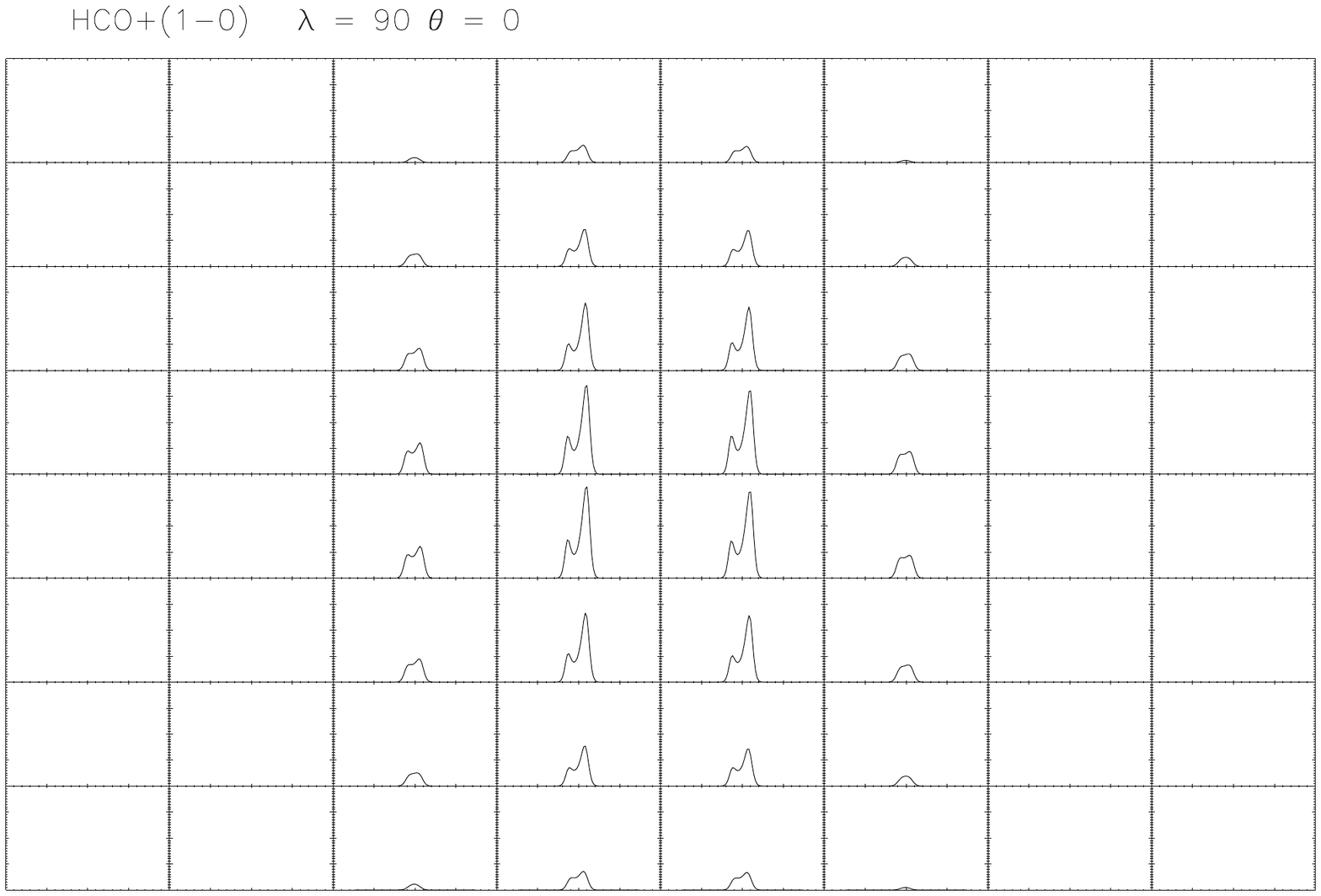}
\end{array}
$
\caption{
({\it Left}) First moment of the simulated HCO$^+$(1-0)
line on a projection along the Z axis. The axes are labeled in arc minutes assuming 
a distance of 250 pc. Velocities in km s$^{-1}$. ({\it Right}) Spectra at intervals
of  0.42 arc minutes. The velocity axis spans a range of -1 to +1 km s$^{-1}$.
The maximum brightness is 3.5 K.
}
\label{fig:2p159_00_90_hcoplus}
\end{figure*}

The velocities are subsonic throughout the simulation resulting in 
asymmetric rather than split spectral line profiles.
The simulated spectral lines seen in an observation on a viewing angle perpendicular to
long axis of the filament
(figure \ref{fig:2p159_00_90_hcoplus})
 are dominated by the outer regions of the
fragment and show the signature of expansion with the red side of the asymmetric
line profile brighter than the blue side. The red asymmetry of the spectral line 
(figure \ref{fig:2p159_00_90_hcoplus} {\it right})
is also seen as red in the first moment map (figure \ref{fig:2p159_00_90_hcoplus} {\it left}).
On a view parallel with the filament, the spectra form a more complex pattern
where the  line profiles are asymmetrical blue and then red in alternating zones. 
Complex patterns of asymmetric spectral lines such as this that
cannot be
created by any combination of rotation and either expansion or contraction and
have been suggested to be an indication of oscillations \citep{Lada2003}.

The simulated spectral line profiles in 
figures \ref{fig:2p159_00_90_hcoplus} ({\it right})
through \ref{fig:3p448_90_00_hcoplus} ({\it right}) may
be compared with actual observations, for example, 
figure 6 of \citet{Lada2003}, figure 3 of \citet{Redman2006},
and figure 2 of \citet{Aguti2007}.

\section{Discussion}

This paper discusses one possible mechanism for the origin of the 
oscillations as a core forms from a filament. Part of the motivation for
this hypothesis is the difficulty of other means. 
Excitation of the oscillations by external forces may not be trivial. 
If two starless cores pass near
each other, the time scale of the encounter needs to be about equal to the internal
sound-crossing time for the core to be tidally excited. This requires a particular
relative velocity. If two cores merge rather than
pass each other, the collision 
has to be slow enough that both cores are not disrupted. The combined mass of 
the merger also has to be low enough that the resulting core is stable against 
gravitational collapse. For example,
\citet{BurkertAlves2009} considered whether the present state of the core B68 were due
to a merger. They concluded that if so, the main core which now appears stable,
would soon collapse with the additional mass of the companion. The problem
with external excitation of the oscillations is that if both 
passing encounters and mergers require special conditions then these mechanisms
would not be suitable to supply oscillation energy to many or most cores.
In contrast, the mechanism for the origin of the oscillations in the origin of the cores
is universal.

We have not addressed the origin of our initial state, the filamentary fragments, or why
that state should be such as to form a stable starless core. 
Filaments and cores are common and intimate in low-mass star-forming regions. Self-gravitating 
filaments are unstable to longitudinal fragmentation on the scale of a Jeans length. This scale
would produce cores with a Jeans mass, similar to the critically stable mass.

With our highly idealized and symmetric initial state, we cannot properly address the
stability of the cores during their formation. The reason is that from our choice of initial
state, the initial longitudinal contraction is destabilizing, similar to the breathing mode in a spherical cloud
defined as the (0,0,0) mode in radial and spherical harmonics . 
In breathing mode oscillations, the inward momentum in the compressive phase acts 
like an additional external pressure and may squeeze an otherwise stable, self-gravitating
cloud past its point of stability and cause it to collapse rather than rebound. 
In contrast, oscillations with higher order modes are stabilizing. We would need more
complex initial states to investigate stability. For example, \citet{Broderick2010} assumed
fully turbulent cores in their analysis of stability, but this already assumes the oscillations 
whose origins we seek to explain.

Nothing requires that all cores form in a stable state. 
For example, star forming regions such as Taurus are defined by a high star formation rate
indicating a large number of unstable cores. Many of these star-forming cores could have been 
unstable from formation. The energy in oscillations may help explain why there are currently
a large number of cores that otherwise appear to be unstable and why the core lifetimes are
statistically longer than their free-fall times.

Magnetic fields, not included in our simulation, could affect the evolution of the cloud
\citep{Hennebelle2003,Galli2005}.
We assume that the magnetic energy is
in equipartition with the kinetic energy. Since most of the energy in observed clouds
is thermal, the magnetic energy should be a minor contributor. If the gas 
velocities in filaments prior to contraction are 
turbulent (subsonic) then the field should also be disordered. 
In this case the effect of the field may be approximated as an additional pressure.
However, if the
field were ordered and anchored on larger scales, the geometry of the initial
filament could be quite different depending on the geometry of the field.

\section{Conclusions}

A hydrodynamic simulation of a contracting filamentary fragment
is able to produce oscillating cores with some complexity
despite the high degree of symmetry in our choice of initial state.
The general agreement between simulated and actual spectral
line observations 
maintains the viability of this mechanism for the origin of the oscillations.

\clearpage

\begin{figure*}
$
\begin{array}{cc}
\includegraphics[trim=0.10in 5.5in 0.8in 1.0in, clip,width=3.25in,angle=180]{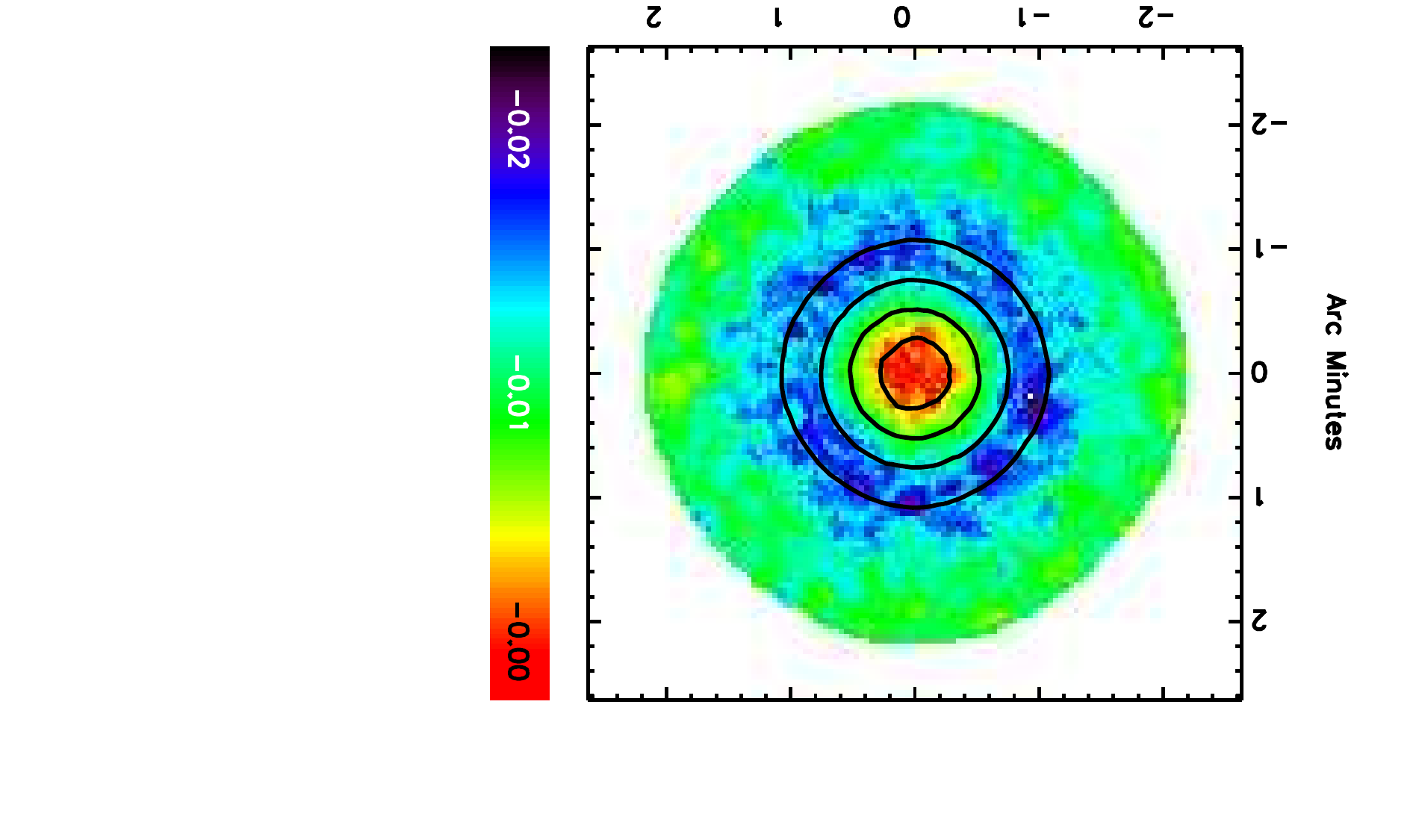}
\includegraphics[trim=0.10in 1.6in 0.8in 0.4in, clip,width=3.25in,angle=180]{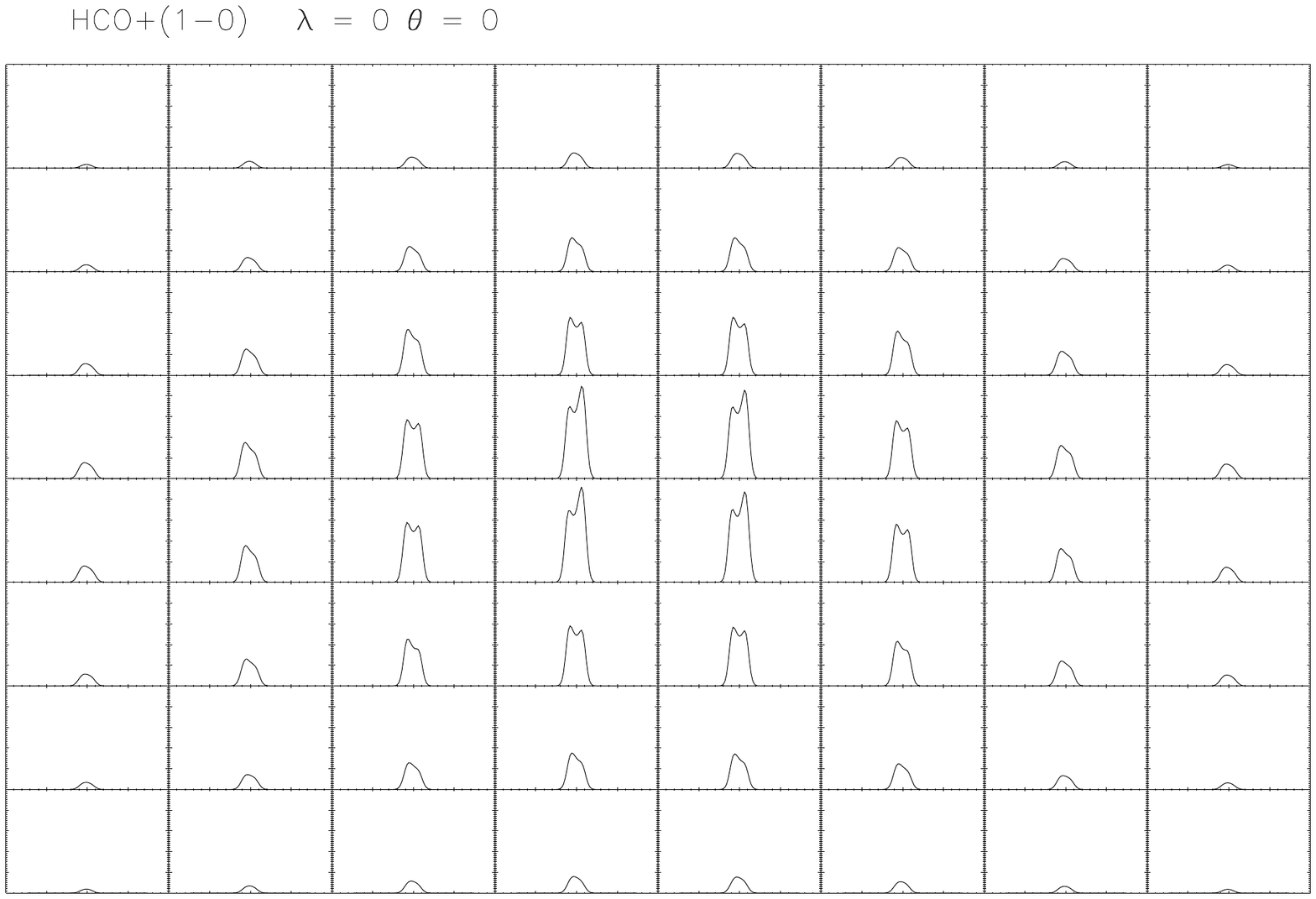}
\end{array}
$
\caption{
Same as figure \ref{fig:2p159_00_90_hcoplus} but on
a projection on the YZ axis. The maximum brightness is 4.6 K.
}
\label{fig:2p159_90_00_hcoplus}
\end{figure*}

\clearpage

\begin{figure*}
$
\begin{array}{cc}
\includegraphics[trim=0.10in 5.0in 0.8in 0.5in, clip,width=3.25in]{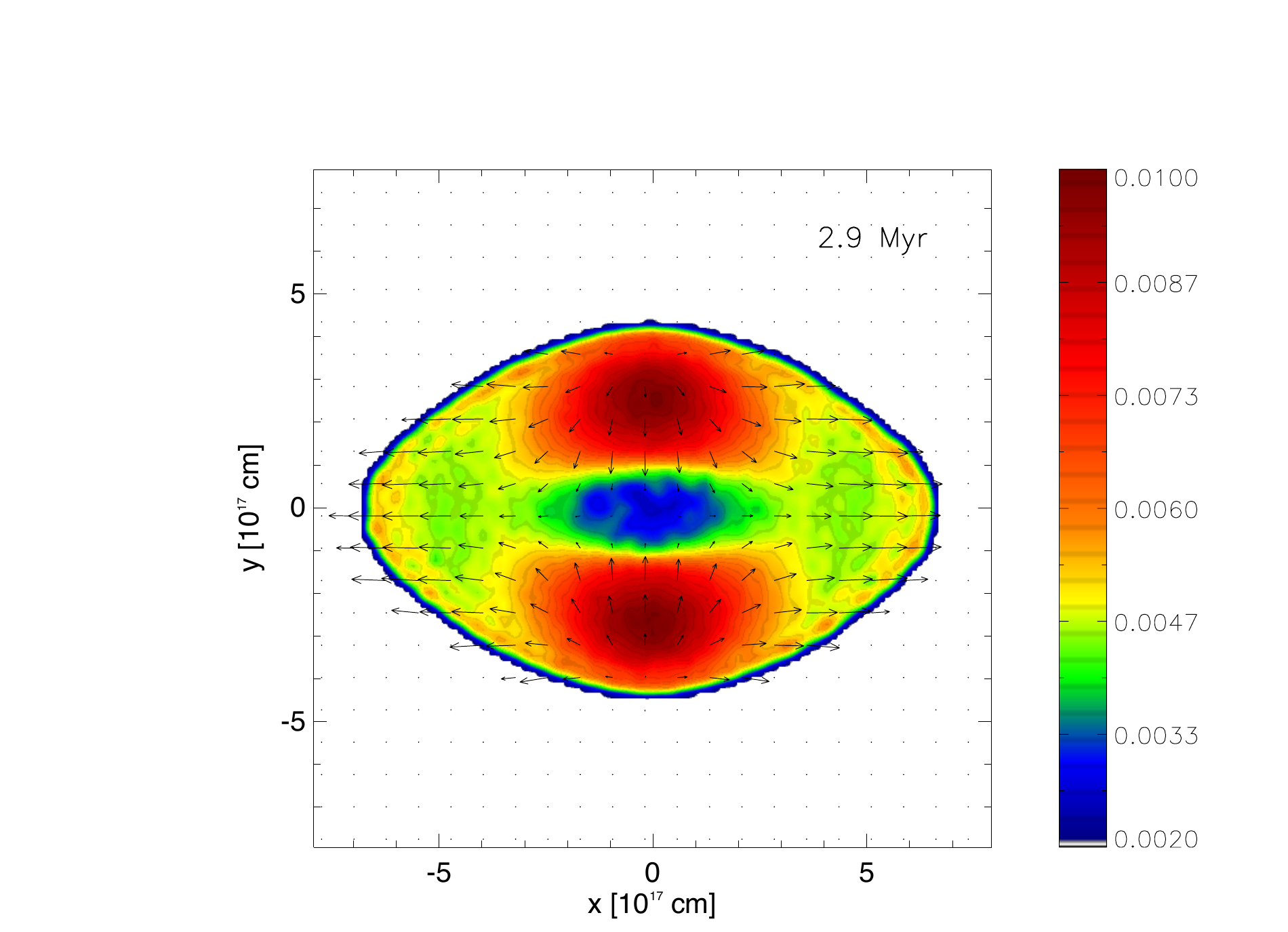}
\includegraphics[trim=0.10in 5.0in 0.8in 0.5in, clip,width=3.25in]{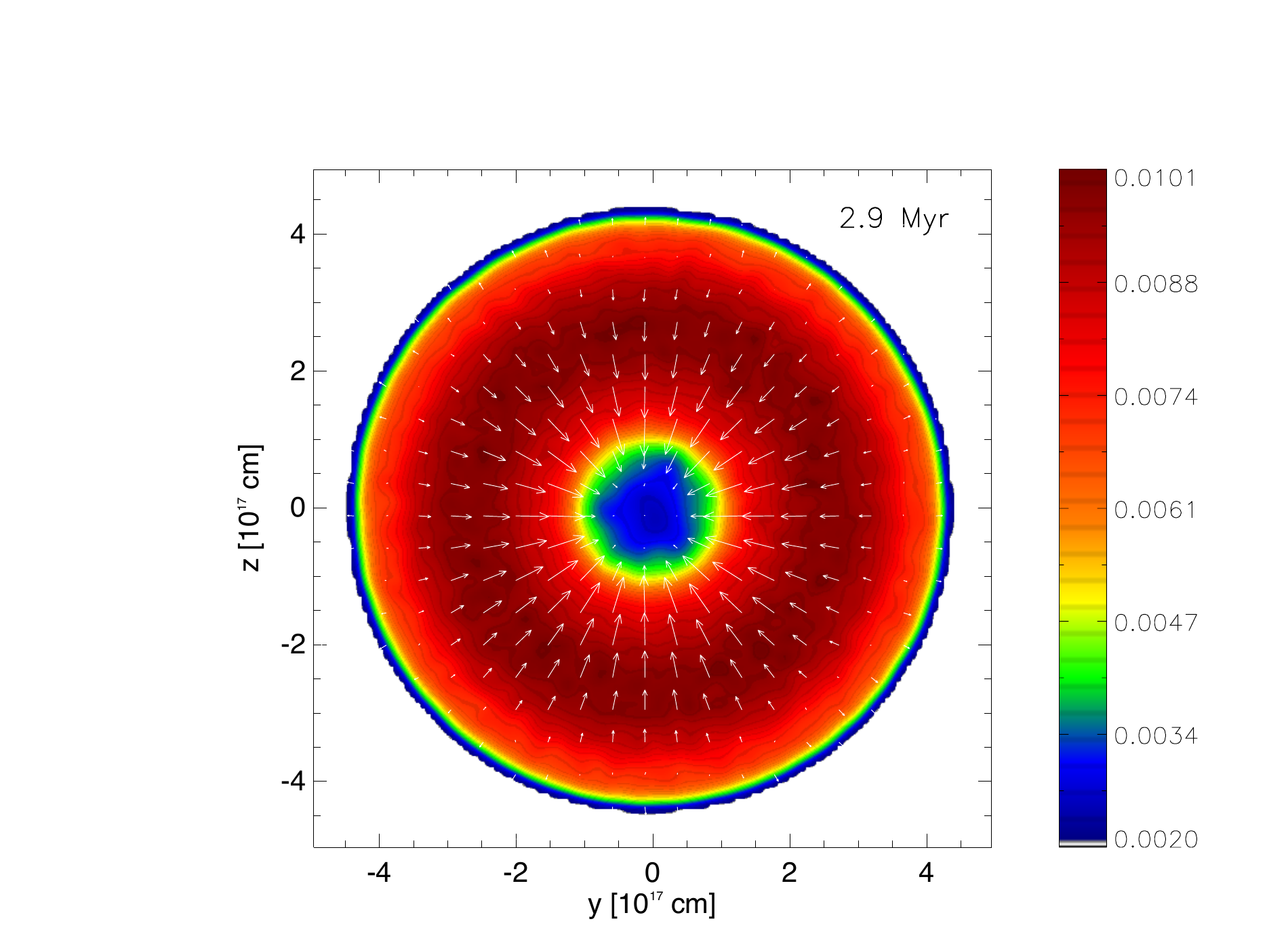}
\end{array}
$
\caption{
Same as figure \ref{fig:figxy-2p159} except for a time of 2.9 Myr. The maximum
velocity is 0.17 km s$^{-1}$.
}
\label{fig:figxy-2p867}
\end{figure*}

\begin{figure*}
$
\begin{array}{cc}
\includegraphics[trim=0.10in 5.5in 0.8in 1.0in, clip,width=3.25in,angle=180]{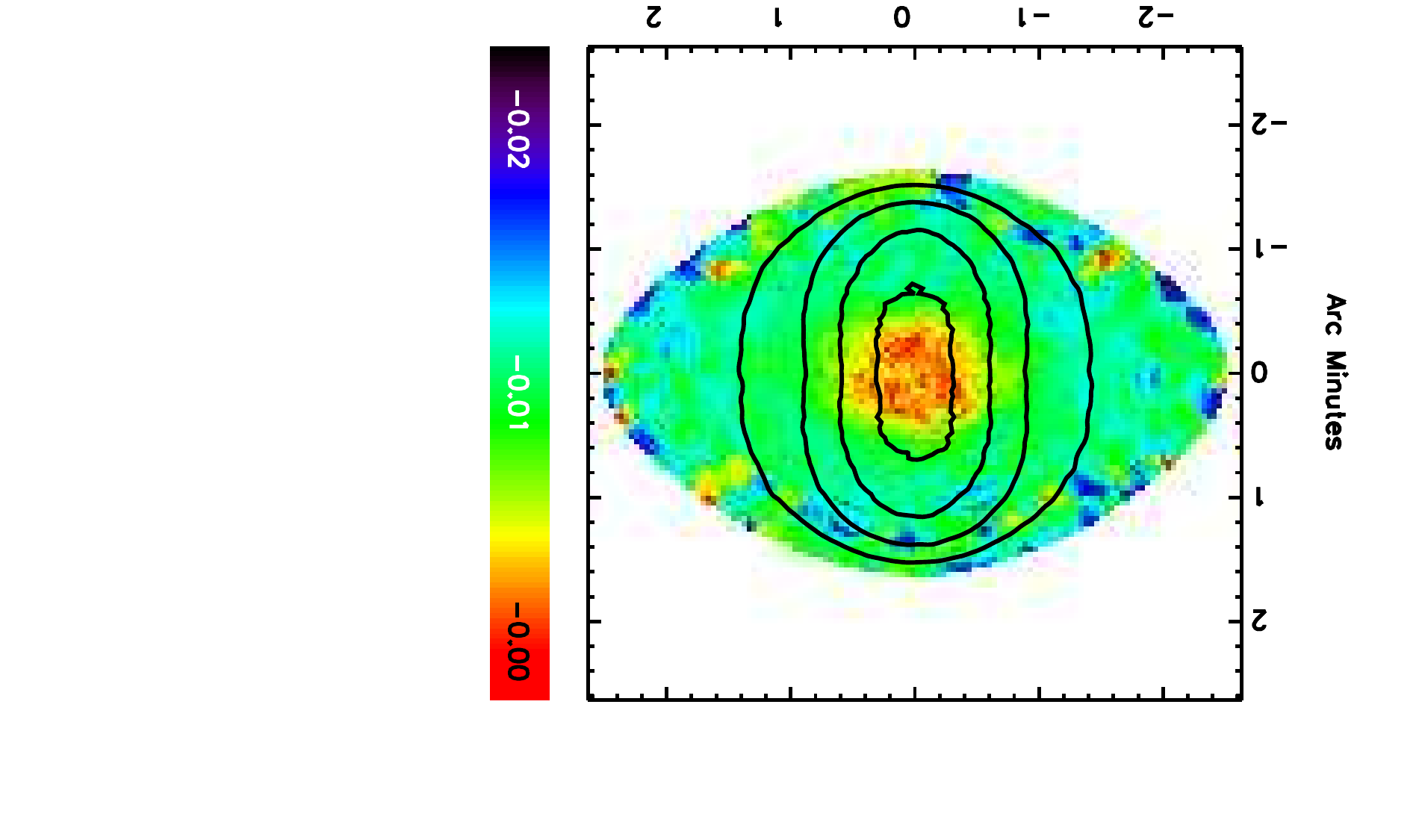}
\includegraphics[trim=0.10in 1.6in 0.8in 1.0in, clip,width=3.25in,angle=180]{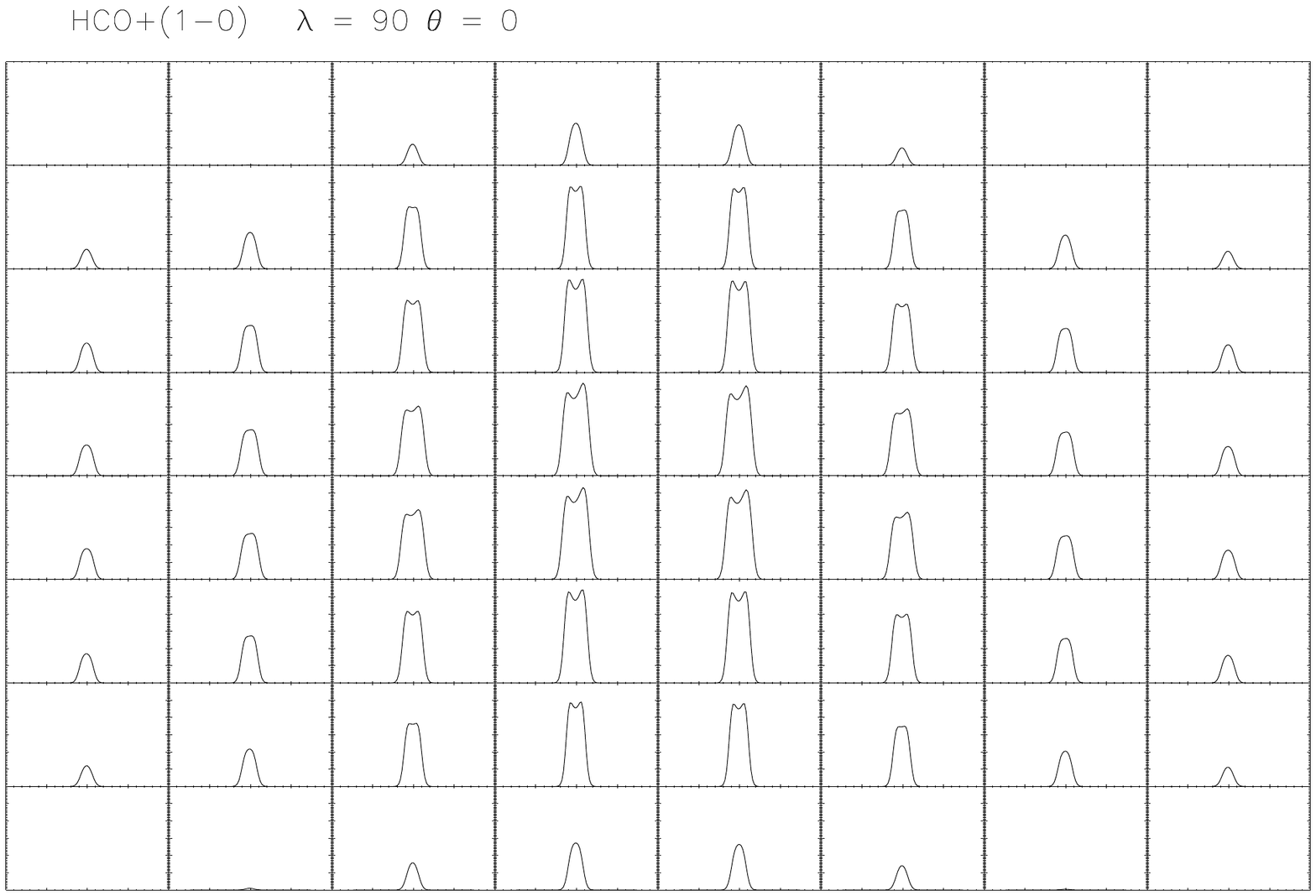}

\end{array}
$
\caption{
Same as figure \ref{fig:2p159_00_90_hcoplus} but for a time of 2.7 Myr. 
The maximum brightness in the spectra is 0.5 K.
}
\label{fig:2p867_00_90_hcoplus}
\end{figure*}

\begin{figure*}
$
\begin{array}{cc}
\includegraphics[trim=0.10in 5.5in 0.8in 1.0in, clip,width=3.25in,angle=180]{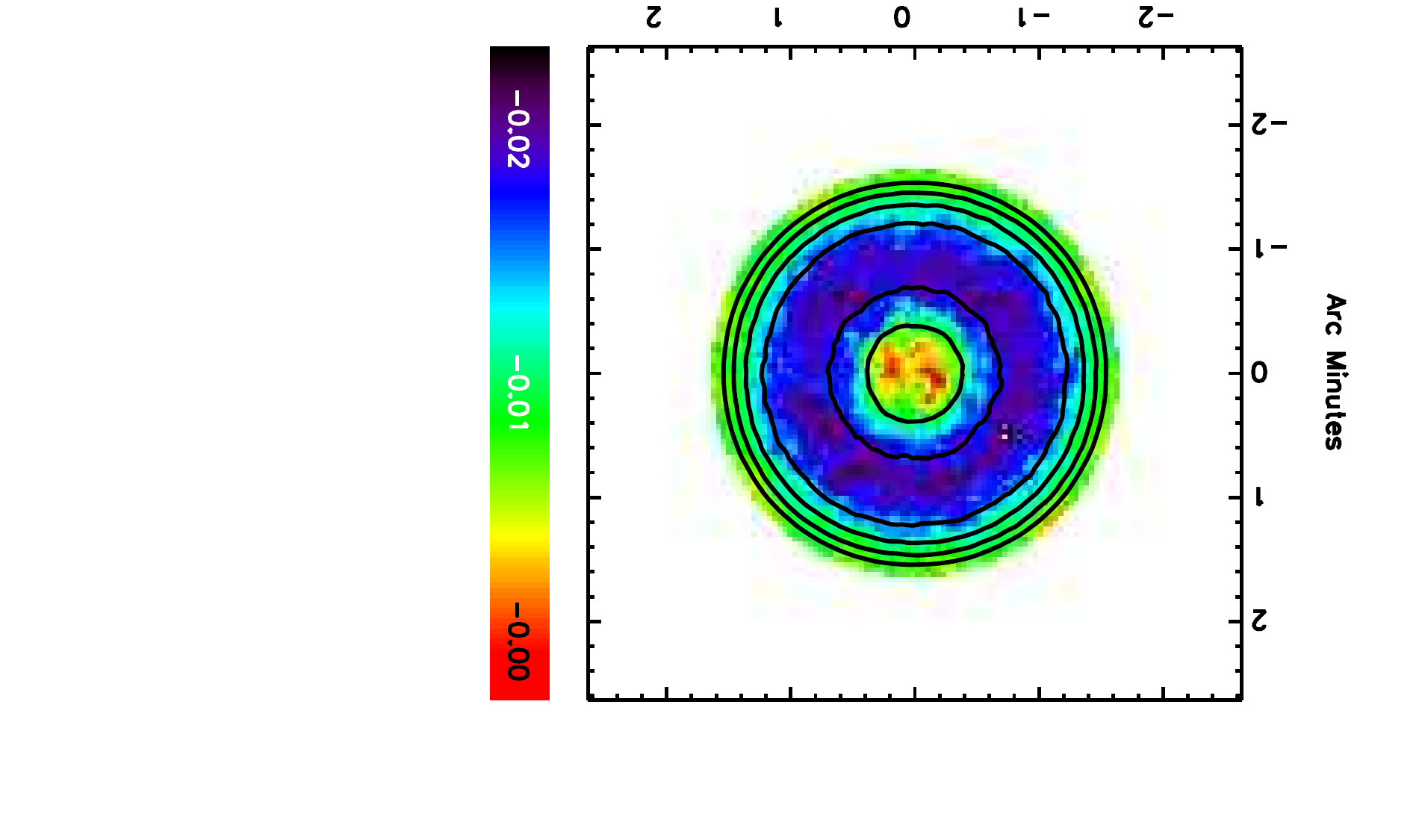}
\includegraphics[trim=0.10in 1.7in 0.8in 1.0in, clip,width=3.25in,angle=180]{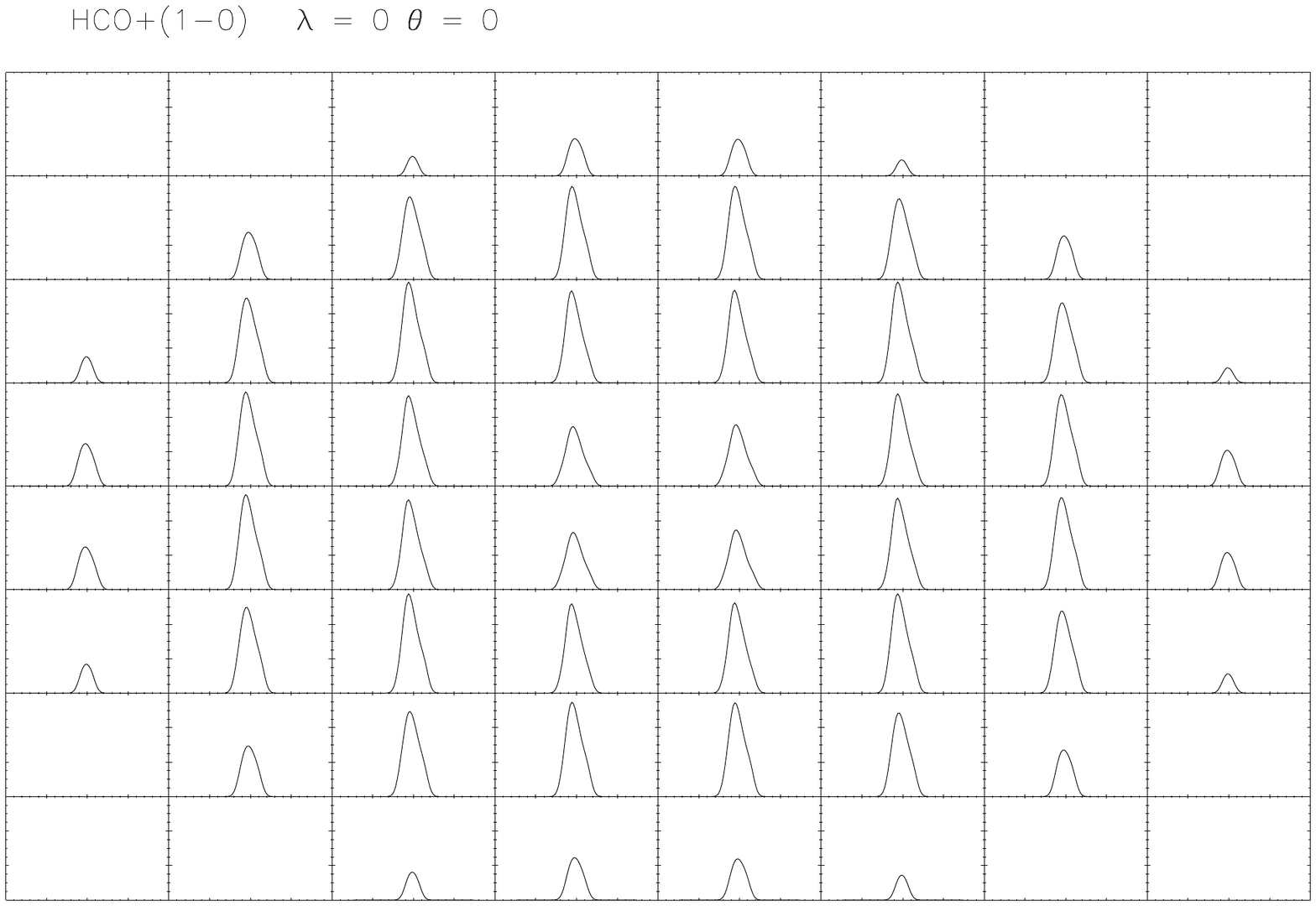}
\end{array}
$
\caption{
Same as figure \ref{fig:2p867_00_90_hcoplus} but on
a projection on the YZ axis. The maximum brightness is 0.6 K
}
\label{fig:2p867_90_00_hcoplus}
\end{figure*}

\clearpage

\begin{figure*}
$
\begin{array}{cc}
\includegraphics[trim=0.10in 5.0in 0.8in 0.5in, clip,width=3.25in]{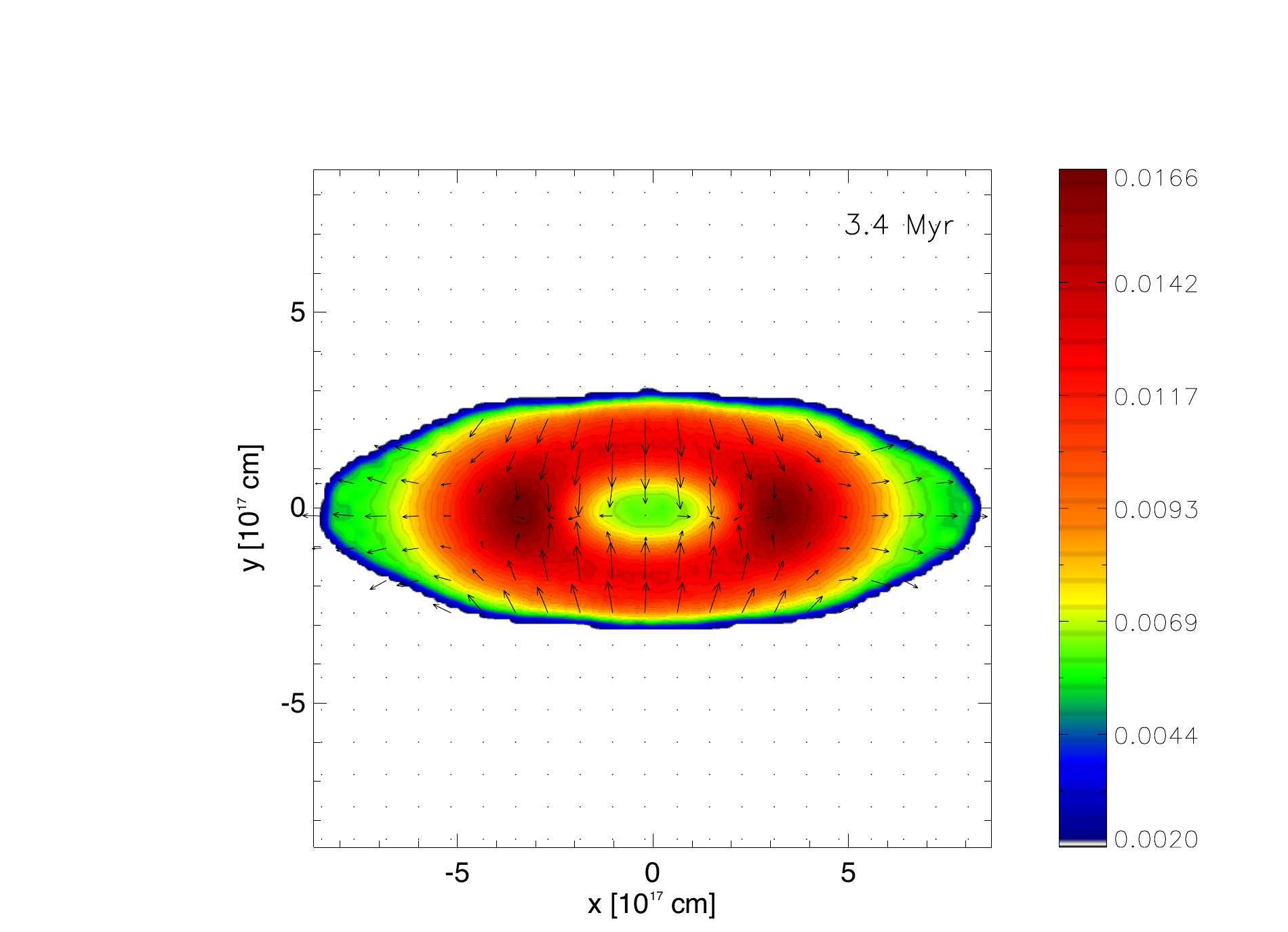}
\includegraphics[trim=0.10in 5.0in 0.8in 0.5in, clip,width=3.25in]{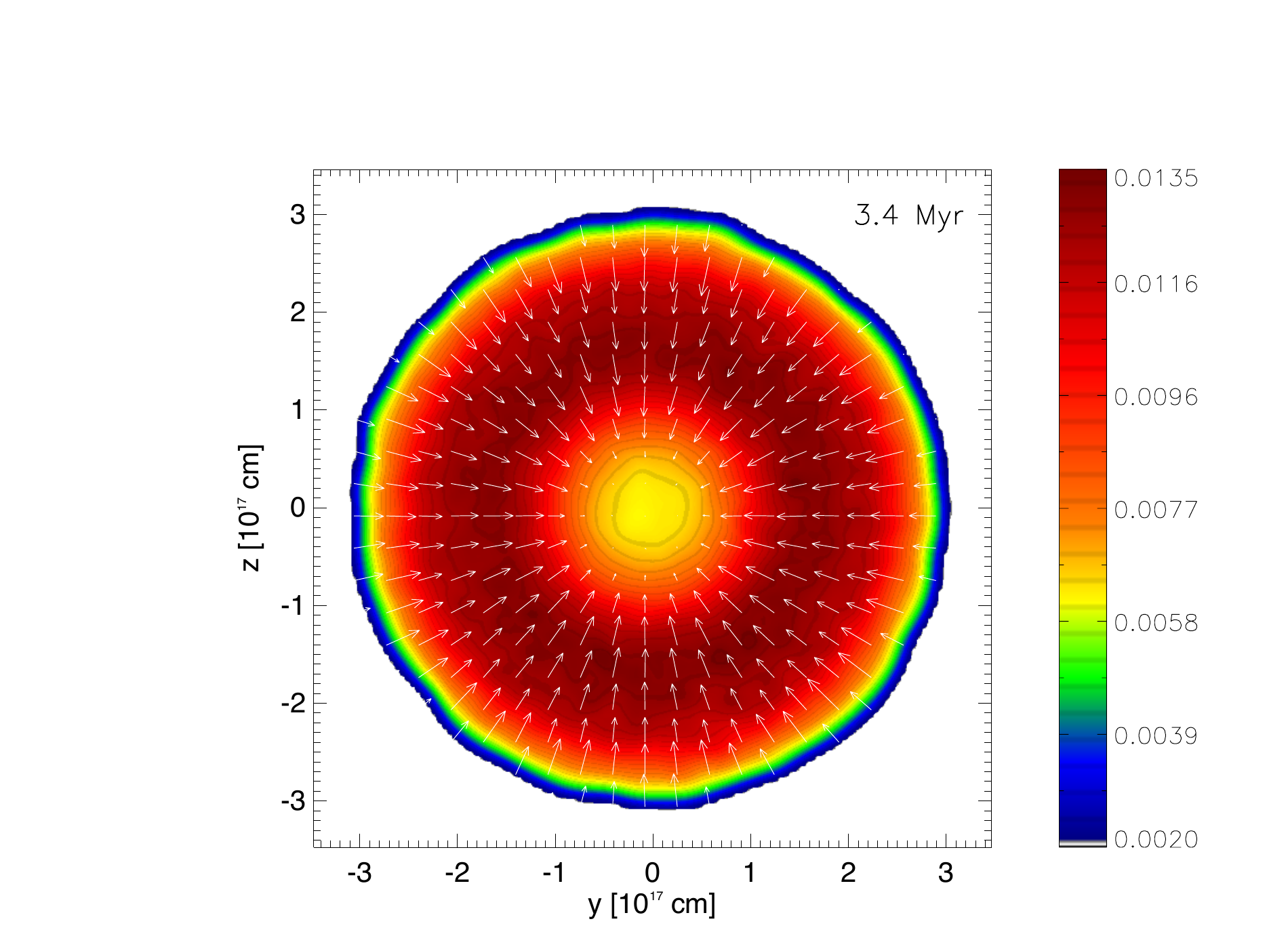}
\end{array}
$
\caption{
(figxy-3p448)
Same as figure \ref{fig:figxy-2p159} except for a time of 3.4 Myr. The maximum
velocity is 0.08 km s$^{-1}$.
}
\label{fig:figxy-3p448}
\end{figure*}

\begin{figure*}
$
\begin{array}{cc}
\includegraphics[trim=0.10in 5.5in 0.8in 1.0in, clip,width=3.25in,angle=180]{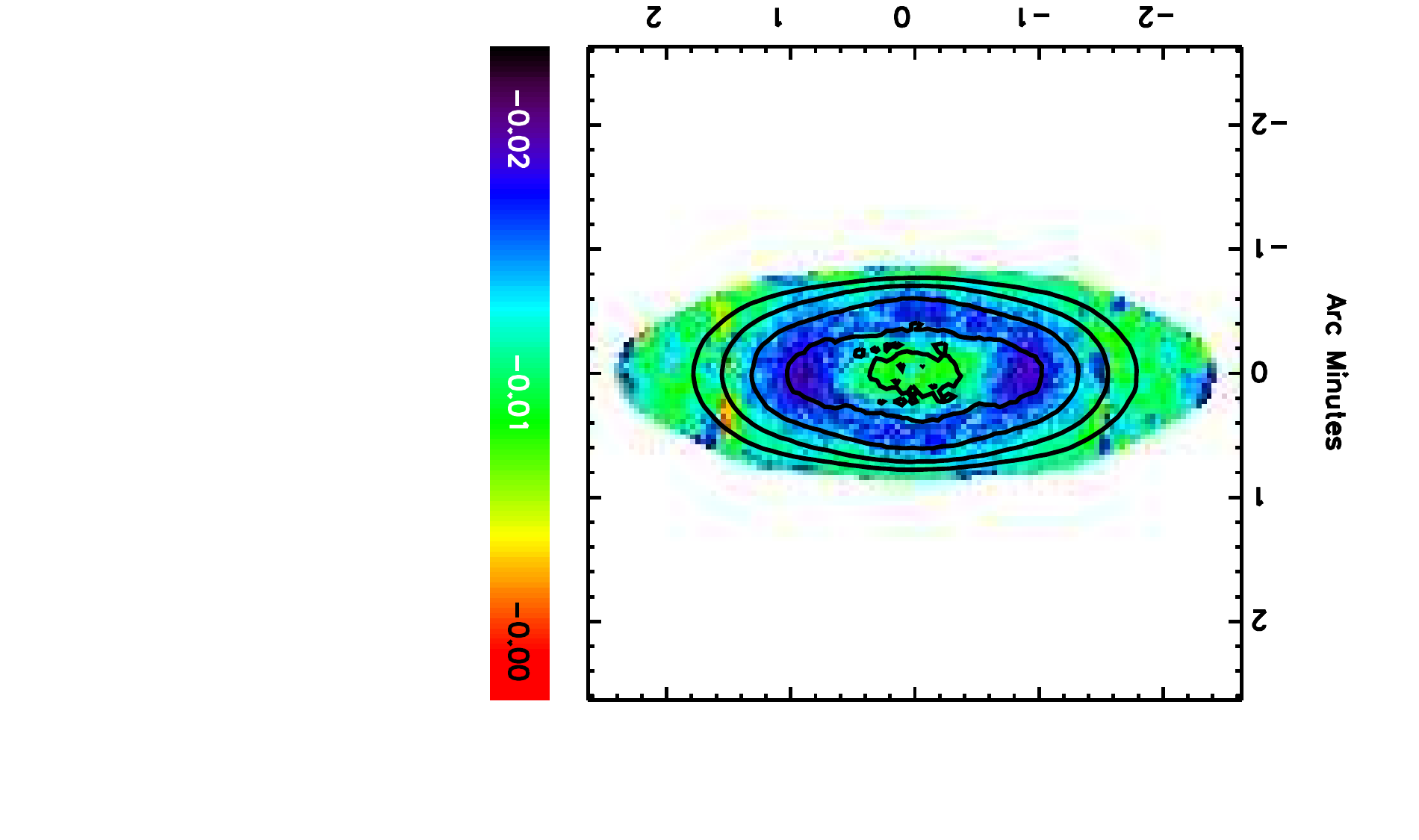}
\includegraphics[trim=0.10in 1.7in 0.8in 1.0in, clip,width=3.25in,angle=180]{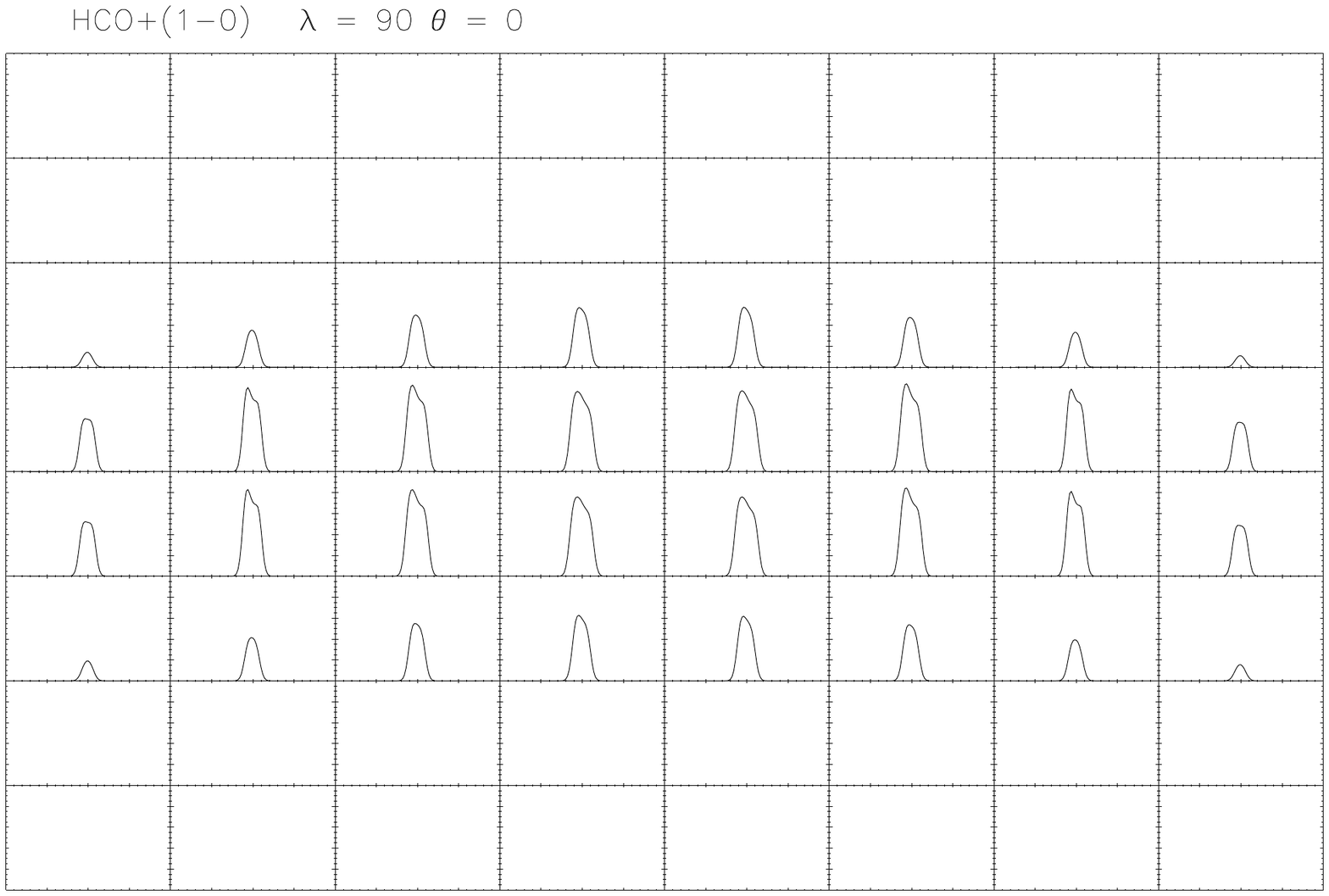}
\end{array}
$
\caption{
Same as figure \ref{fig:2p159_00_90_hcoplus} but for a time of 3.5 Myr. 
The maximum brightness in the spectra is 0.8 K.
}
\label{fig:3p448_00_90_hcoplus}
\end{figure*}

\begin{figure*}
$
\begin{array}{cc}
\includegraphics[trim=0.10in 5.5in 0.8in 1.0in, clip,width=3.25in,angle=180]{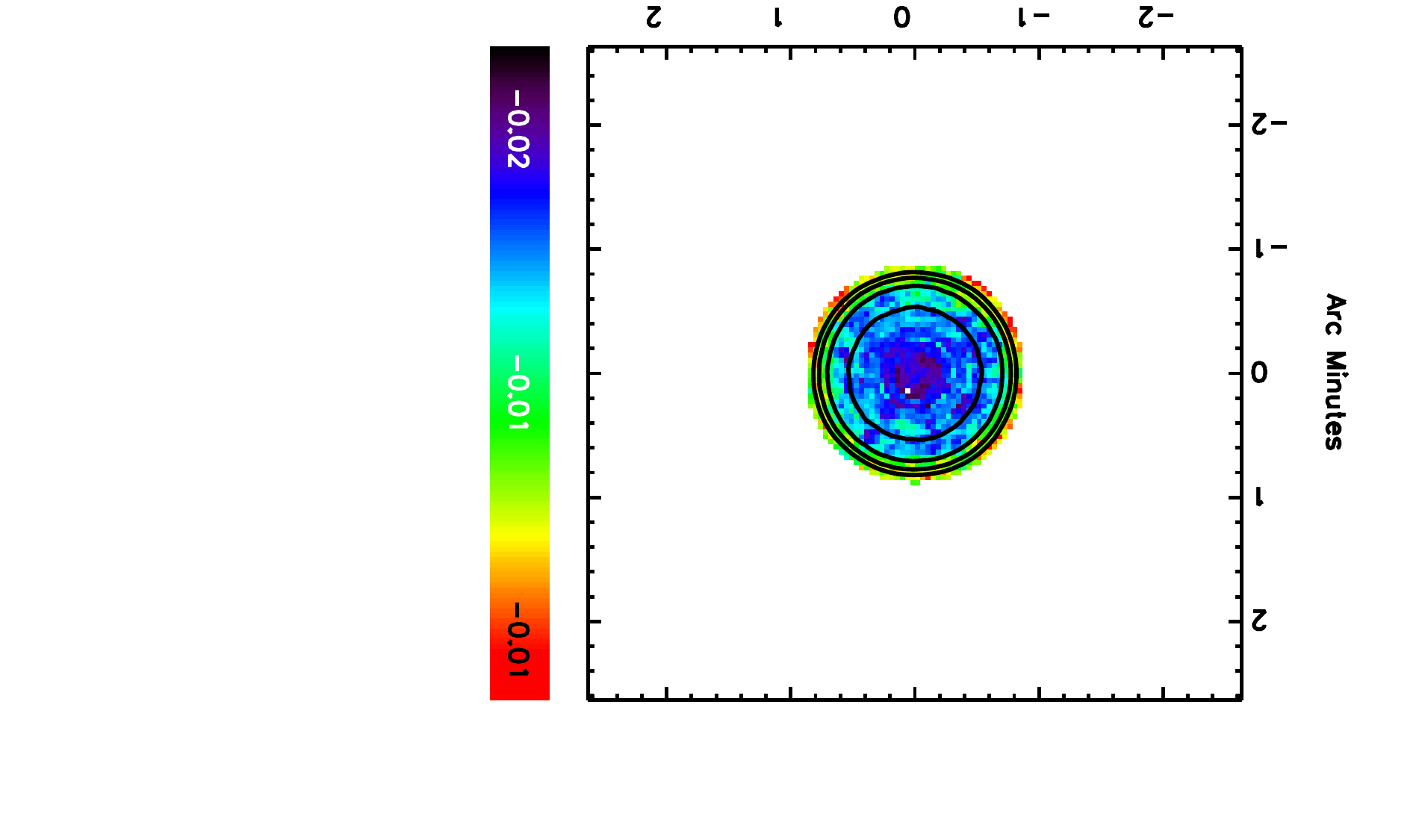}
\includegraphics[trim=0.10in 1.7in 0.8in 1.0in, clip,width=3.25in,angle=180]{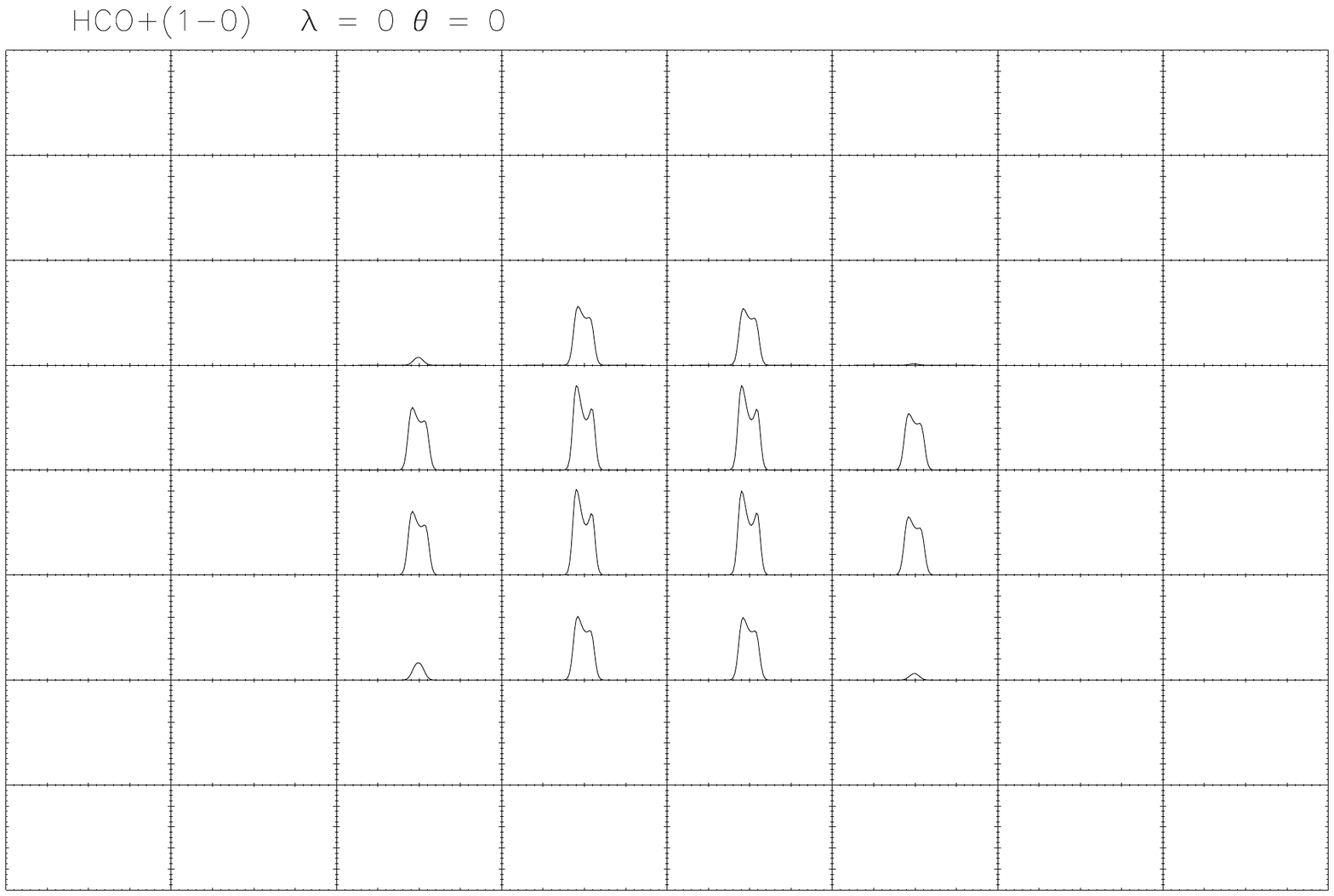}

\end{array}
$
\caption{
Same as figure \ref{fig:3p448_00_90_hcoplus} but on
a projection on the YZ axis. The maximum brightness is 0.8 K.
}
\label{fig:3p448_90_00_hcoplus}
\end{figure*}

\clearpage

\begin{figure}
\includegraphics[trim=3.7in 3.8in 0.2in 0.2in, clip,width=3.25in,angle=0,scale=0.8]{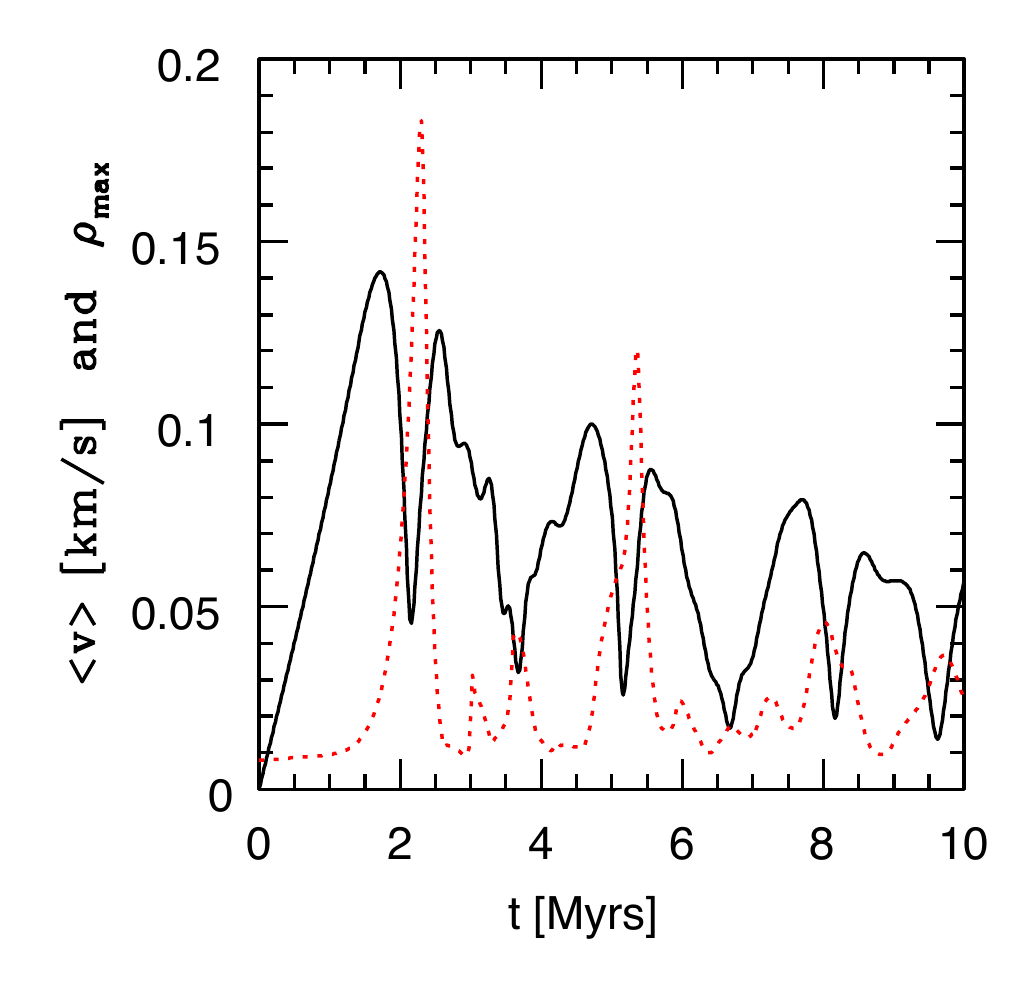}
\caption{
Average velocity ({\it black solid line})and maximum density ({\it red dotted line})
as a function of time. The average velocity is defined by the 
total kinetic energy, $E_{kin}$  divided by the total mass, $M$,  
$v=\sqrt{2*E_{kin}/M}$. The units of density are $2\times 10^{-18}$ g cm$^{-3}$.
The figure shows the 2 periods of oscillation, longitudinal and radial with the
highest densities occurring at the point of maximum compression when the velocities
are low and reversing.
}
\label{fig:evolution}
\end{figure}

\noindent{\bf Acknowledgements} A.B. thanks the Harvard-Smithsonian  Center for Astrophysics for their hospitality during multiple visits.
The research of A.B. is supported by the priority program 1573 "Physics of the Interstellar Medium"
of the German Science Foundation.

\bibliography{kb4}

\end{document}